\documentclass{mem}
\usepackage{natbib}
\usepackage{balance}
\usepackage{txfonts}
\usepackage{graphicx}
\idline{75}{283}

\begin{document}

\title{GRID-Launcher v.1.0}

\subtitle{}

\author{N. \,Deniskina\inst{1}, M. \,Brescia\inst{2}, S. \,Cavuoti\inst{1}, G. \,d'Angelo\inst{1}, O. \,Laurino\inst{1},\\
G. \,Longo\inst{1}$^,$\inst{2}}

    \offprints{N. Deniskina}

\institute{
Department of Physical Sciences \- University Federico II, Naples, via Cinthia 6 
\- 80131 Naples, Italy, 
\email{natalia.deniskina@unina.it}
\and 
INAF - Osservatorio Astronomico di Capodimonte, via Moiariello 16, 80131, Napoli, Italy}

\authorrunning{N. Deniskina et al.: }

\titlerunning{GRID-Launcher v.1.0}

\abstract{GRID-launcher-1.0 was built within the VO-Tech framework, as a software interface between 
the UK-ASTROGRID and a generic GRID infrastructures in order to allow any ASTROGRID user to launch 
on the GRID computing intensive tasks from the  ASTROGRID Workbench or Desktop.
Even though of general application, so far the Grid-Launcher has been tested on a few 
selected softwares (VONeural-MLP, VONeural-SVM, Sextractor and SWARP) and on the SCOPE-GRID.

}

\maketitle{}

\section{Introduction}

The main goal of the International Virtual Observatory \cite{IVOA} infrastructure is 
to provide the community at large with an easy and user friendly access to astronomical data,
software tools, and computing facilities. 
While the first two parts of the task, namely the federation and fusion of heterogeneous 
data archives and the implementation of flexible data reduction and data analysis 
tools have been widely addressed and, at least in their fundamental aspects, solved, the 
possibility to access large distributed computing facilities to perform computing 
intensive tasks has not yet been satisfactorily answered.
The main reasons for this delay being mainly two: the need to match the GRID security 
requirements and the lack of homogeneity in the definitions of the storage space.
The first issue can be easily explained: most users of a specific Virtual Organization (VO)
do not possess the personal certificates which are requested to access the GRID or, even
when they do have a personal certificate, the computing GRID which they need does not recognize 
their own certification authority. 
The second issue arises instead from the fact that the data usually reside locally or in a remote 
storage space which is not seen by the selected GRID as a storage element (SE). 

The lack of an user friendly access to the GRID is a main obstacles against the use of computing 
intensive data mining methods and tools such as, for instance, Support Vector Machines \cite{cavuoti} 
or Probabilistic Principal Surfaces \cite{dabrusco} on massive data sets (MDS). 
Therefore, in the framework of the VONeural project, which aims at implementing a package of data 
mining routines capable to work in a distributed computing environment and on large data sets of 
high dimensionality, we have implemented and tested a general purpose interface between the 
UK-ASTROGRID \cite{astrogrid} and the GRID which solves part of the above quoted problems.
We wish to stress that the followed approach is quite general and that GRID-Launcher
can be easily adapted to other Virtual Organizations and to any GRID. For testing
we used the GRID-SCOPE \cite{scope} which is part of the recently funded  
GRID infrastructure for Southern Italy.

\section{GRID-launcher}

As already mentioned, an ASTROGRID user is recognized by the UAS (User Authentication Service) of ASTROGRID but,
often does not possess a personal GRID certificate. 
As it is the case for most virtual organizations, the ASTROGRID and GRID authentication procedures 
are incompatible due to a number of factors: a) different CAs (certification authority);
b) the lack of compatibility between the ASTROGRID and GRID certificates and PKI policy. 
The EGEE-II boards are currently discussing how to allow the use of specific services by means of 
application certificates (or ''robot'' certificates) but, at the time GRID-Launcher 1.0 was developed 
this option had not yet be introduced and the first version discussed here needs therefore to be regarded 
as a test version of a more flexible tool (GRID-Launcher 2.0) which will be made available as soon as the 
new standards will be introduced.

In other words, the version 1.0 relies on a personal GRID certificate (signed by the INFN-GRID CA) which
is recognized by the SCOPE-GRID used for testing.

\begin{figure}[t!]
\resizebox{\hsize}{!}{\includegraphics[clip=true]{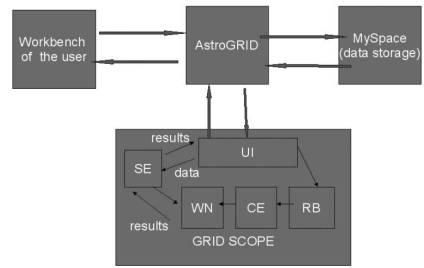}}
\caption{\footnotesize Workflow. UI: user interface; 
RB: resource broker; SE: storage element; CE: computing element; 
WN: working node.} \label{figa}
\end{figure}

\begin{figure}[t!]
\resizebox{\hsize}{!}{\includegraphics[clip=true]{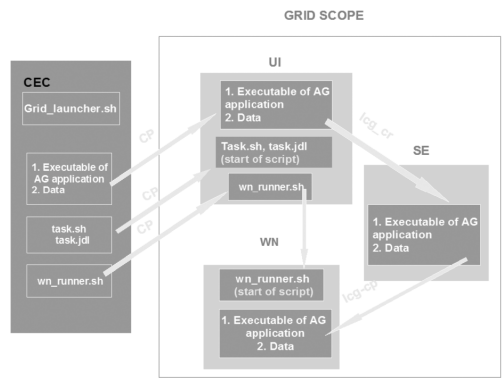}}
\caption{\footnotesize The workflow from ASTROGRID to the WN of the
GRID. The task.sh, wn\_runner.sh and task.jdl - are automatically generated 
by the GRID launcher. } \label{figb}
\end{figure}

\begin{figure}[t!]
\resizebox{\hsize}{!}{\includegraphics[clip=true]{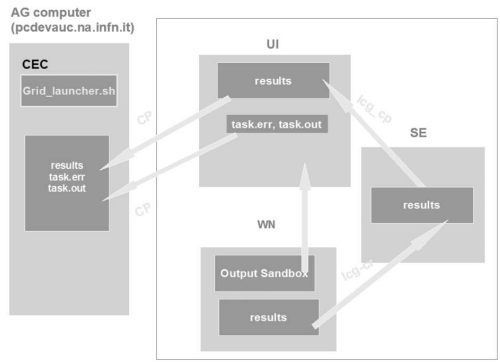}}
\caption{\footnotesize The workflow from the WN to ASTROGRID.
The task.err and task.out files are automatically generated by the 
GRID. } \label{figc}
\end{figure}

\noindent In the figures 1-3, we give a schematic representation of the way GRID-Launcher 
works: 
(Fig.\ref{figa})

\begin{enumerate}
\item Grid-launcher:
\begin{itemize}
\item takes the user input from the User Interface of ASTROGRID (Workbench);
\item collects all files, tabs and programs needed;
\item generates automatically three scripts:task.sh, task.jdl and wn\_runner.sh 
to be executed on the GRID;
\item wraps them in an archive and sends it to the Scope-GRID UI. (The authentication on the GRID 
is done by public keys exchange).
\end{itemize}
\item The GRID User Interface (UI) receives the data and the JDL program,
unpacks them, copies the data to the SE, copies wn\_runner.sh to the WN's,
starts task.sh and task.jdl. (Fig. \ref{figb})
\item wn\_runner.sh starts on the WNs, takes the data from SE, starts the
executable of the application, and puts the results on the SE. 
The GRID generates automatically two output files task.err and task.out and
sends them to the UI using the Output SandBox.
\item GRID-launcher periodically checks the status of job and when it ends, it 
moves the results from the UI to the ASTROGRID machine. (Fig. \ref{figc})
\item GRID-launcher receives the data archive, unpacks it and puts
the results into the storage of ASTROGRID (VO-Space).
\end{enumerate}

SO far, GRID-Launcher has been implemented and tested on:
\begin{itemize}
\item VO-Neural-MLP: a Multi Layer Perceptron Neural Network used for 
supervised clustering and classification \cite{skordovski,voneural_2};
\item VO-Neural-SVM: Support Vector Machines for supervised clustering \cite{cavuoti,voneural_2};
\item Sextractor: for the extraction of object-catalogues from astronomical images \cite{SEx},
\item SWarp: an application that resamples and coadds FITS images using any arbitrary astrometric 
projection defined in the WCS standard\cite{swarp}.
\end{itemize}

\section{Future developments}
In the version 2.0, we plan:
\begin{itemize}
\item to insert a "switch" which will check whether the user has a GRID certificate (and in this case the 
task will start on the GRID of the user), or has not (in this case the program will be launched as a service on 
the SCOPE-GRID);
\item to render the program capable to launch any program (and not only applications registered in the 
ASTROGRID) matching some trivial specifications;
\item to offer the possibility to use GRID-Launcher as a subtask inside the ASTROGRID workflows and, finally,
\item to unify the VO Space with the SE of the GRID.
\end{itemize}

Additional information about "GRID-launcher" can be found at \cite{voneural}.

\bibliographystyle{aa}

\begin{thebibliography}{}

\bibitem[Bertin \& Arnouts, 1996]{SEx} Bertin E. \& Arnouts S., 1996, A\& AS, 117, 393

\bibitem[Brescia et al. 2008]{voneural_2} Brescia M., et al., 2008, these proceedings.

\bibitem[Cavuoti 2008]{cavuoti} S. Cavuoti, Laurea Thesis, 2008, University Federico II in Napoli

\bibitem[D'Abrusco et al. 2008]{dabrusco} D'Abrusco R., Longo G., Walton N., 2008, astro-ph/0805.0156v1
\bibitem[Skordovski 2008]{skordovski} B. Skordovski, Laurea Thesis, 2008, University Federico II in Napoli

\bibitem[URL.1]{astrogrid} URL.1: http://www.astrogrid.uk/
\bibitem[URL.2]{IVOA} URL.2: http://www.ivoa.org/
\bibitem[URL.3]{scope} URL.3: http://scope.unina.it/
\bibitem[URL.4]{swarp} URL.4: SWarp User manual , E. Bertin at: http://terapix.iap.fr/rubrique.php?id\_rubrique=49
\bibitem[URL.5]{voneural} URL.5: http://voneural. na.infn.it/


\end{thebibliography}

\end{document}